Alessandro Frigeri, Gisella Speranza

# "Eppur si muove"
## Software libero e ricerca riproducibile

> "Il gioco della scienza è, in linea di principio, senza fine. Chi, un bel giorno, decide che le asserzioni scientifiche non hanno più bisogno di nessun controllo, e si possono ritenere verificate definitivamente, si ritira dal gioco".
>
> Karl Popper

Supponiamo che Robinson Crusoe fosse dotato di una geniale intelligenza scientifica e che avesse potuto disporre, nella sua isola deserta, delle tecnologie più avanzate. Seppure fosse giunto, da solo e disponendo di un tempo illimitato, a dimostrare le più complesse leggi scientifiche che noi abbiamo scoperto nei secoli, non potrebbe essere comunque considerato uno scienziato. Popper, ne *La società aperta e i suoi nemici*, si serve di questa celebre immagine per chiarire il carattere necessariamente pubblico della scienza.

Secondo il filosofo tedesco, l'insieme di sistemi teorici che ipotizziamo che Crusoe abbia sviluppato – per quanto li possiamo immaginare coerenti, complessi e ottenuti seguendo il metodo sperimentale – non costituiscono un sapere scientifico, perché nel suo mondo non esiste nessun altro, eccetto lui stesso, che possa mettere alla prova le sue teorie. Non possono essere concepiti altri sistemi teorici con cui il suo possa entrare in conflitto, o risultati scientifici differenti che inducano il nostro scienziato a rivedere e migliorare i suoi procedimenti. La scienza dunque costituisce un sapere collettivo, che non può prescindere dalla condivisione tra membri di una comunità. Nel caso di Crusoe, non si può parlare di scienza perché nella sua isola viene a mancare un elemento fondamentale del metodo: la riproducibilità dell'esperimento finalizzata al controllo della teoria. Curiosamente, il paradigma della "scienza crusoniana", usato per rappresentare l'impossibilità di un sapere scientifico al singolare, si rivela incredibilmente efficace per descrivere alcune pratiche diffuse negli attuali ambienti della ricerca.

L'introduzione delle tecnologie dell'informazione nella scienza ha cambiato radicalmente le sue forme e i suoi strumenti. Ha permesso di elaborare grandi quantità di dati attraverso complesse procedure numeriche e in tempi relativamente brevi, ottenendo dei risultati prima impossibili da pianificare. In questo contesto rinnovato, spesso lo scienziato non è più colui che esegue gli esperimenti in laboratorio o che ne analizza i risultati, ma colui che sa far funzionare un computer, a cui affida il compito di risolvere un problema. Apparentemente, con l'introduzione dei calcolatori, insieme all'immagine dello scienziato, cambiano radicalmente le dinamiche della ricerca. Se, però, proviamo a confrontare i due modelli – quello, per così dire, classico con la ricerca informatizzata – ci rendiamo conto che sono sostanzialmente sovrapponibili, poiché le procedure informatiche, che costituiscono il software, rappresentano l'implementazione della formula matematica della legge scientifica.

Secondo il modello di Popper, il controllo di una teoria è attuato riproducendo le esperienze fattuali ad essa correlate e valutando la corrispondenza dei risultati ottenuti con l'asserto teorico. Se questi risultano conformi alla teoria, accrescono la probabilità della sua correttezza, mentre basta che un solo risultato la falsifichi, per accertarne l'inattendibilità. Perché una ricerca possa essere controllata e risponda alle esigenze del metodo scientifico deve essere garantita l'accessibilità ai dati e alla loro procedura di elaborazione. Nel caso in cui questa sia costituita da un software – invece che da procedimenti matematici – è lo stesso software che necessita di controllo.

Va precisato che un programma informatico è un insieme organizzato di istruzioni elementari impartite ad una macchina tramite un *codice sorgente,* che viene da questa tradotto in un *codice binario.* Mentre il primo è scritto in un linguaggio di programmazione *intermedio* tra l'uomo e la macchina, il secondo è il risultato della sua *compilazione* in un linguaggio eseguibile dal computer ma non intelligibile da un programmatore, il *linguaggio macchina*. Allo stesso modo, i dati digitali che vengono elaborati e prodotti dal software sono una sequenza binaria codificata attraverso regole dettate da un *formato dati*, che ne permette l'interpretazione.

Affinché venga garantito il principio di intersoggettività del metodo scientifico – che afferma che "nella scienza possono essere introdotte soltanto asserzioni tali da poter essere controllate intersoggettivamente" (K. Popper, *Logica della scoperta scientifica*, Einaudi, Torino 1970) – è necessario pubblicare, insieme ai programmi e ai dati, anche codice sorgente e formato dati. Soltanto in questo modo una ricerca potrà dirsi scientifica.

Inizialmente i calcolatori venivano forniti alle istituzioni di ricerca completi di programmi e codici sorgente, e i programmatori erano gli stessi ricercatori che collaboravano per sviluppare software sempre più efficienti e funzionali ai loro scopi scientifici. A partire dagli anni '80 si affermò gradualmente una consuetudine che prevedeva che il software non fosse più disponibile liberamente, e che venisse gestito soltanto dall'istituzione che ne commissionava la compilazione. Nacque così il cosiddetto *software proprietario*, che inibiva agli utenti l'accesso al codice sorgente, affidando lo sviluppo, il controllo e le modifiche del programma ad un gruppo ristretto di specialisti.

Nel 1984 Richard Stallman, ricercatore del laboratorio di intelligenza artificiale del Massachusetts Institute of Technology di Boston, notò che il proprio lavoro era fortemente limitato dall'introduzione del software proprietario: un programma che non funzionava dovere non poteva essere modificato dagli stessi ricercatori, a meno che questi non si rendessero disponibili a firmare accordi di segretezza con le ditte che possedevano il controllo dello sviluppo del software, piegandosi a comportamenti di dubbia correttezza professionale nei confronti dei propri colleghi. Stallman si licenziò dal proprio lavoro per cercare di risolvere quello che aveva avvertito come un problema limitante per la propria professione. Era sua ambizione creare un *sistema operativo* che fornisse programmi per gli utenti e garantisse la piena accessibilità agli strumenti di sviluppo dei software che lo costituivano, offrendo la possibilità di compilare nuovi programmi. L'idea di software libero formalizzata da Stallman – basata sulla possibilità di utilizzare il software per qualsiasi scopo, di poterlo copiare, di poterne studiare il codice sorgente e di poterlo migliorare – si adatta perfettamente alle esigenze di pubblicità e condivisione della ricerca scientifica. È in grado, infatti, di innescare un meccanismo virtuoso di controlli, modifiche ed evoluzioni dei programmi, che, superando di gran lunga i limiti ristretti dell'istituzione proprietaria, potenzialmente assume le dimensioni illimitate della comunità scientifica mondiale, grazie anche alle possibilità di diffusione globale e istantanea delle informazioni offerte da internet.

Proprio mentre il sistema operativo GNU di Stallman cercava un "nucleo" abbastanza efficiente su cui funzionare, nel 1991 è nato Linux, creato da uno studente di ventuno anni dell'Università di Helsinki, Linus Torvalds, che lo aveva sviluppato per la sua tesi e lo mise in rete sotto forma di software libero, proprio per avere suggerimenti per migliorarlo. In questo modo tutti potevano lavorarci, accedendo al codice sorgente, e pubblicare nuovamente i risultati in rete con tutte le informazioni necessarie a un ulteriore controllo. Così, dal 1991, il sistema operativo GNU/Linux, permette a tutti di poter utilizzare, condividere e sviluppare software libero.

I principi di Stallman disegnano una comunità scientifica aperta, in cui la libera circolazione dei risultati della ricerca è condivisa da una pluralità di soggetti, che garantiscono il controllo delle nuove esperienze acquisite, lo scambio di conoscenze e quindi l'accrescimento del sapere. In questo modo si salvaguarda il momento dialogico e critico, parte integrante della genesi di un qualsiasi asserto scientifico, che non è mai da intendere come definitivo e immutabile, ma anzi, pretende di essere controllato, integrato, modificato o, se necessario, rifiutato.

Una comunità scientifica di questo tipo, basata sulla condivisione delle conoscenze e sulla controllabilità dei risultati, si pone in continuità con l'idea popperiana di una scienza che "non persegue mai lo scopo illusorio di rendere le sue risposte definitive", ma che, libera da pregiudizi di ogni tipo, progredisce mirando allo "scopo infinito, e tuttavia raggiungibile, di scoprire problemi sempre nuovi, più generali e più profondi, e di sottoporre le sue risposte, date in via di tentativo, a controlli sempre rinnovati e sempre più rigorosi" (K. Popper, *Logica della scoperta scientifica,* Einaudi, Torino 1970). In questo senso, l'errore, che è un possibile risultato del controllo di una teoria, è il punto di partenza per un suo miglioramento o per il suo abbandono definitivo, e costituisce quindi un'occasione di progresso.

Il software proprietario, al contrario, si basa sulla segretezza del codice sorgente e del formato dei dati, considerati una proprietà esclusiva della ditta produttrice. In questo caso, la comunità scientifica che mette in atto il meccanismo di controllo e di scoperta degli errori si restringe all'ambito chiuso di una sola azienda, che non diffonde all'esterno l'implementazione degli algoritmi e delle procedure, riducendone l'attendibilità e ostacolandone il miglioramento. Si chiede così al resto della comunità scientifica di riporre la fiducia in procedure non riproducibili, introducendo un principio di autorità da sempre estraneo al metodo scientifico.

La pratica proprietaria della segretezza del codice sorgente e del formato dati è motivata dalle stesse dinamiche di rivalità commerciale per cui da sempre viene tenuta segreta la ricetta della Coca Cola. L'atto di sigillare le procedure informatiche si fonda su un'idea di software inteso come bene materiale, assimilabile a un qualsiasi prodotto commerciale. È bene chiarire le conseguenze di questo cambiamento radicale delle dinamiche proprie della comunità scientifica. In realtà la funzione che il software svolge nella ricerca è del tutto assimilabile al ruolo della successione di formule atte a risolvere un problema scientifico. Entrambi, infatti, costituiscono la procedura di elaborazione di dati sperimentali secondo le determinate variabili che interagiscono nel fenomeno da studiare. Il software, quindi, non è un bene materiale, come non lo è il teorema di Pitagora. È evidente che una simile confusione tra un bene intellettuale e un bene materiale autorizza a pratiche che contaminano la diffusione del sapere nella comunità scientifica, così come è stata teorizzata dalle sue prime formulazioni fino a Popper. Se Pitagora non avesse scritto del suo teorema, o se Euclide non ne avesse pubblicato la dimostrazione della procedura matematica, o fosse stato impedito in qualche modo l'uso della sua formula, nessuno avrebbe potuto utilizzare per qualsiasi altra applicazione la nota procedura per calcolare l'ipotenusa di un triangolo rettangolo a partire dai due cateti. Sarebbe stato necessario inventare dei metodi alternativi per eseguire lo stesso calcolo, invece di usarlo direttamente nello sviluppo di teorie più complesse.

Il software, così come la formula scientifica, è un bene culturale. Questo, però, non vuol dire che non possa essere commercializzato, come accade per ogni altra produzione intellettuale. L'elemento nuovo di questo commercio, piuttosto, sta nel fatto che l'acquisto del software proprietario ne concede solo l'utilizzo e non la conoscenza degli elementi che lo costituiscono, negando così la possibilità che l'intera procedura, o alcuni dei suoi elementi, possano essere utilizzati per sviluppare altri algoritmi. Il software open-source non intende sovvertire – come spesso si vuole far credere – le regole della proprietà intellettuale previste nella nostra società. Non ci sono, infatti, elementi di incompatibilità tra l'idea di una proprietà intellettuale e la libera accessibilità al sapere. Il copyright, come tutela dell'autore del software, è assolutamente rispettato e formalizzato in licenze d'uso di pieno valore legale (la più nota delle quali è la General Public License). La differenza è che gli strumenti della proprietà intellettuale sono interpretati dagli sviluppatori di software libero come mezzi per tutelare sia l'utente che lo sviluppatore.

In realtà le consuetudini proprietarie ricalcano più la logica del brevetto che quella del diritto d'autore. E non è un caso che la pratica di brevettare software si stia già diffondendo negli Stati Uniti, mentre sulla delicata questione della sua effettiva validità legale si è aperto un ampio dibattito in sede di parlamento europeo. Il brevetto è da sempre uno strumento di parziale salvaguardia di un qualsiasi prodotto tecnologico dalla competizione commerciale. Applicare questo meccanismo di protezione ad un software vuol dire considerarlo uno strumento della ricerca, invece che una sua parte integrante, con tutte le gravi conseguenze di metodo che questo equivoco comporta quando si tratta di programmi finalizzati alla ricerca scientifica.

Molti istituti di ricerca pubblici e privati usano GNU/Linux, sviluppano software libero e promuovono l'accessibilità ai dati attraverso formati aperti, applicando le pratiche di cui una sana produttività scientifica ha bisogno. La scuola di ricerca sull'esplorazione sismica dell'Università di Stanford, fondata dal prof. Claerbout, ha da sempre promosso una ricerca riproducibile e ha trovato recentemente nel software libero l'ambiente ideale per svilupparla. Nonostante ci siano vari celebri esempi di questo tipo, fino a che nella ricerca coesisteranno entrambe le tipologie di software, la comunità scientifica non sarà un sistema del tutto aperto e la scienza sarà incapace di progredire pienamente secondo le proprie dinamiche di sviluppo, che si nutrono di una critica costante ad ogni nuova acquisizione. Inoltre i gruppi di ricerca che si impegnano nella produzione e nel miglioramento di software, e che ne pubblicano i dati saranno fortemente penalizzati, se il loro lavoro continuerà a non essere valutato come parte integrante della ricerca stessa.

Sulla base di queste osservazioni, si può sostenere che l'utilizzo del software proprietario nella comunità scientifica costituisce una minaccia per alcuni fondamentali principi del metodo. Al contrario, c'è un'evidente analogia tra l'idea di comunità scientifica descritta da Popper e le dinamiche di diffusione delle procedure di ricerca condivise dagli sviluppatori del software libero.

La storia della scienza è costellata di episodi più o meno celebri in cui i principi del metodo sono stati scavalcati da fattori esterni che, in modo diverso, hanno sostituito all'atteggiamento critico un principio di autorità, limitandone i caratteri di apertura e libertà. Ci sembra che oggi la pratica del software proprietario, applicata in ambito scientifico, possa costituire un nuovo "nemico" di quella che secondo Popper è la "società aperta" per eccellenza, cioè la comunità scientifica. Usando un parallelo non così fuori luogo come può sembrare, potremmo dire che quello che per Galileo ha rappresentato il sistema tradizionale di Aristotele e Tolomeo, per la scienza di oggi rischia di diventare l'utilizzo del software proprietario nella ricerca.

*Alessandro Frigeri, Gisella Speranza*

**Per saperne di più:**

K. Popper, *La società aperta e i suoi nemici*, Armando Editore, Roma 2004;
K. Popper, *La logica della scoperta scientifica*, Einaudi, Torino 1970;
http://softwarelibero.it, sito dell'Associazione per il Software libero;
http://www.fsf.org, sito della Free Software Foundation creata da Richard Stallman;
http://www.gnu.org/, sito del progetto GNU;
http://www.linux.it/Vademecum#Legenda_dei_termini, glossario di termini legati al software libero, parte del vademecum della Italian Linux Society.